\documentstyle[editedvolume,psfig]{crckapb}

\begin{opening}
\title{Final performances of the X--ray mirrors
of the JET--X telescope}

\author{E. PORETTI}
\author{S. CAMPANA}
\author{O. CITTERIO}
\author{P. CONCONI}
\author{M. GHIGO}
\author{F. MAZZOLENI}
\author{G. TAGLIAFERRI}
\institute{Osservatorio Astronomico di Brera\\
Via E. Bianchi, 46 -- 23807 Merate, Italy}
\author{G. CUSUMANO}
\author{G. LA ROSA}
\institute{Istituto di Fisica Cosmica CNR\\
Via U. La Malfa, 153 -- 90146 Palermo, Italy}
\author{W. BR\"AUNINGER}
\author{W. BURKERT}
\institute{Max Planck Institut\\
Giessenbachstrasse -- 85748 Garching, Germany}
\author{C.M. CASTELLI}
\author{R. WILLINGALE}
\institute{Dept. Physics and Astronomy, Leicester University\\
University Road -- Leicester, United Kingdom}

\end{opening}

\runningtitle{JET--X Telescopes}

\begin{document}

The Joint European X--ray Telescope (JET--X) is one of the core
scientific instruments of the SPECTRUM RONTGEN-$\gamma$ astrophysics
mission.
The project is a collaboration of British, Italian and Russian
consortia, with the participation of the Max Planck Institut
(Germany).
JET--X was designed to study the emission from X--ray sources in the band of
0.3--10 keV. Citterio et al. (1996 and references therein) describe its 
structure, composed by two identical and coaligned Wolter I telescopes.
Focal plane imaging is provided by cooled
X--ray sensitive CCD detectors which combine high spatial resolution
with good spectral resolution, including coverage of the iron line complex
around 7 keV at a resolution of $\Delta E/E\sim2\%$.

To measure the effective area of the telescopes, the detectors (PSPC or CCD)
were either directly exposed to the incoming X--ray beam or put in
the focal plane collecting the photons reflected by the telescopes. 
The ratio between the two exposures gave the effective area values. 
The mean values obtained by using the two flight models were 161 
cm$^2$ at 1.5 keV and 69 cm$^2$ at 8 keV, respectively. These values
match very well the theoretical ones (see Figure 3 in Citterio et al. 1996).
Another test, concerning the evaluation of the angular resolution, was
carried out using a CCD detector in the focal plane; after half the exposure
time the detector was moved by a distance equivalent to $20"$.
The two sources are well resolved on the resulting CCD image,
confirming the excellent angular resolution achieved by JET--X
(see Figure 2 in Citterio et al. 1996).

Crowded field regions and extended X--ray emitting sources represent
key targets for the JET--X telescopes. Thus, it is required that
the Half Energy Width (HEW) remains good also at large off--axis angles.
To illustrate the JET--X capabilities, we plot in Fig. 1 the HEW
of one full telescope as a function of the off-axis angle, 
as measured during the end-to-end test at the Panter facility in M\"unchen.
As it can be seen, up to $12'$ at 1.5 keV the HEW is below $\sim 20"$
(filled circles) and at 8.1 keV it is almost constant at a level 
of $\sim 23"$ (open circles). Beyond this off-axis angle, the mirror
figuring errors dominate, producing almost the same HEW at all energies.
The best HEW at 1.5 keV is obtained at $\sim 8'$ off-axis, due to a small
displacement in the JET--X focus, introduced to improve the off-axis response.
As a general result, the end--to--end calibration tests provided a 
successful confirmation of all the JET--X performances.
\begin{figure}
\psfig{figure=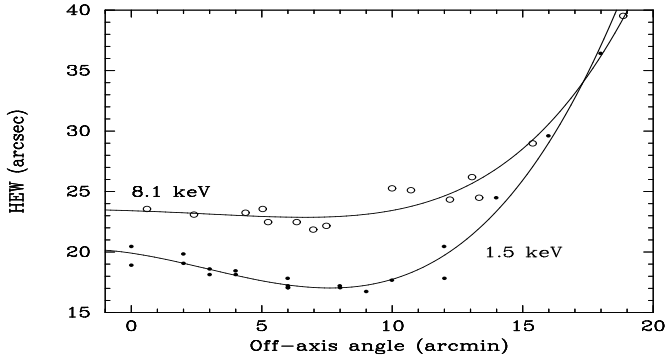,width=6.5truecm,height=6.5truecm}
\caption{The results of the end--to--end tests at the Panter facility 
confirm that the required HEW is achieved also at large off--axis angles.}
\end{figure}

\end{document}